# HIGH DIMENSIONAL PROCESS MONITORING USING ROBUST SPARSE PROBABILISTIC PRINCIPAL COMPONENT ANALYSIS

Mohammad Nabhan, Yajun Mei and Jianjun Shi

H. Milton Stewart School of Industrial and Systems Engineering, Georgia Institute of Technology, Atlanta, GA 30332

**ABSTRACT**

High dimensional data has introduced challenges that are difficult to address when attempting to implement classical approaches of statistical process control. This has made it a topic of interest for research due in recent years. However, in many cases, data sets have underlying structures, such as in advanced manufacturing systems. If extracted correctly, efficient methods for process control can be developed. This paper proposes a robust sparse dimensionality reduction approach for correlated high-dimensional process monitoring to address the aforementioned issues. The developed monitoring technique uses robust sparse probabilistic PCA to reduce the dimensionality of the data stream while retaining interpretability. The proposed methodology utilizes Bayesian variational inference to obtain the estimates of a probabilistic representation of PCA. Simulation studies were conducted to verify the efficacy of the proposed methodology. Furthermore, we conducted a case study for change detection for in-line Raman spectroscopy to validate the efficiency of our proposed method in a practical scenario.

# 1. INTRODUCTION

With the deployment of large numbers of sensors and the wide use of imagery in monitoring, the monitoring of high dimensional data streams that result from these systems has gained a lot of interest in recent years. Traditional statistical process monitoring procedures may fall short in these data rich environments. A popular approach to address this issue is to reduce the dimension of the available data streams. Principal Component Analysis (PCA) is a ubiquitously used dimension reduction technique (Jolliffe 2011). PCA is a method that projects a set of observed variables onto a significantly lower dimensional subspace spanned by directions referred to as principal components. The resulting projection points are commonly referred to as PC scores. However, PCA has been shown to produce extremely inconsistent estimates in high dimensional settings, when the low dimensional space is sparse (Ma 2013). Not to mention the inherent issues that arise from poor interpretability as the estimated principal components are linear combinations of all the data streams (Archambeau and Bach 2009, Guan and Dy 2009).

An example of such data rich environments is the continuous production of carbon nanotubes (CNTs) buckypaper. A recent development in the inspection process utilizes in-line Raman spectroscopy (Yue et al. 2018). The ability to monitor the manufacturing process in real time is critical to scale it up while meeting high quality standards. However, it is challenging to detect changes in the data collected from this procedure. This is because the obtained profiles are high dimensional with specific segments where peaks occur as illustrated in Figure 1. In addition to the sparsity of these features, the noise is complex with signal-dependent properties and may be confused with defects (Yue et al. 2017).

In this article, we propose a new method that combines the two properties of sparsity and robustness within a probabilistic framework. A probabilistic approach provides a direct platform for change detection and fault diagnosis, which will be used to address the monitoring challenges of the motivating Buckypaper production process. In addition, a probabilistic method has potential in addressing incomplete or missing data via conditional probability densities as well as extensions to mixture models (Archambeau et al. 2008, Tipping and Bishop 1999a). However, the latter properties will not be discussed in this study as they deserve to be autonomously investigated.

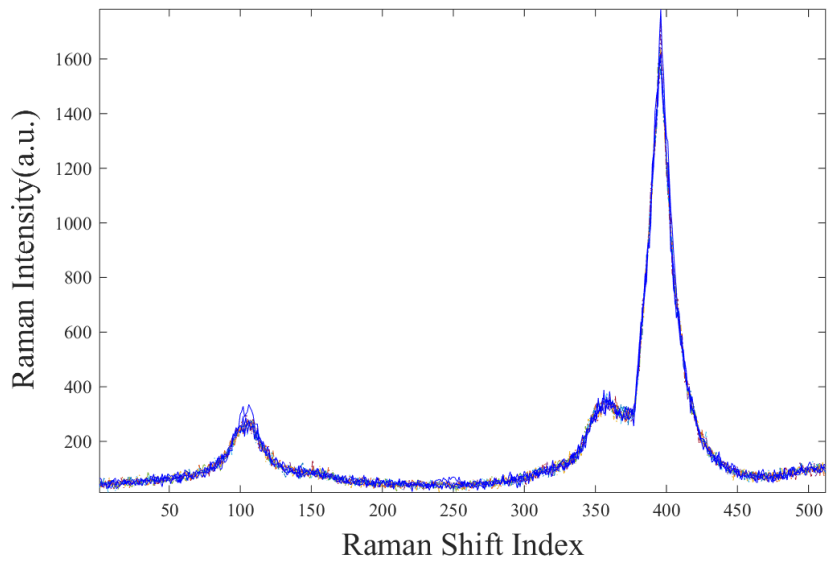

Figure 1 Illustration of the Raman spectra data

Since all statistical conclusions are based in a probabilistic space, a Mahalanobis type distance measure is more appropriate than a uniform distance measures (e.g. the Euclidean distance) for extracting the principal components (Kim and Lee 2003). Several probabilistic PCA variations have been proposed for process monitoring including, some are robust and others are sparse but to the best of our knowledge none that combine both (Zeng et al. 2017, Zhu et al. 2014, Chen et al. 2009). In the case study and the monitoring portion of the simulation experiments, we evaluate the performance of our robust and sparse probabilistic approach for process monitoring by comparing it with other probabilistic methods that lack one or both properties.

Preceding work on the problem of obtaining both sparse and robust principal components include (Hubert et al. 2016, Croux et al. 2013), which will be used as benchmarks in the principal component extraction portion of the simulated experiments. The aforementioned work combine the two properties by developing sparse modifications of existing robust formulations of PCA, namely projection pursuit PCA (PP-PCA) (Croux and Ruiz-Gazen 2005, Li and Chen 1985) and ROBPCA (Hubert et al. 2005). While the aforementioned methods may be effective for the objective of extracting robust and sparse principal components, they are not performed in a probabilistic framework and are not intended for process monitoring.

The remainder of this paper is organized as follows: In Section 2, we provide a brief review of relevant topics in the literature followed by a more detailed overview of dimension reduction methods dealing with sparsity and/or robustness. Next, in Section 3, we illustrate in detail our proposed dimension reduction methodology as well as the monitoring strategy. Section 4 demonstrates the effectiveness of our proposed sampling strategy in virtually simulated scenarios. Furthermore, section 5 presents a case study on change detection for in-line Raman spectroscopy. We then finally conclude the article with a discussion of the major findings of our proposed monitoring scheme.

## 2. LITERATURE REVIEW

This section is split into two subsections. The first subsection 2.1 provides a review of classical PCA and various methodologies used to improve robustness and induce sparsity to the original formulation of PCA. The second subsection 2.2 presents a brief overview of the basic approach for probabilistic principal component analysis (PPCA) and its robust and sparse variations. Finally, subsection 2.3 provides a brief overview of Bayesian variational inference. This progression will lay the necessary foundation necessary to facilitate the discussion of our proposed robust sparse probabilistic PCA method in section 2.3.

## 2.1 Classical, robust and sparse PCA

Classical PCA is a dimension reduction technique that projects a set of observed data $x \in \mathbb{R}^p$ onto a subspace of latent (feature) variables $z \in \mathbb{R}^q$, where the dimension of the latent variables is significantly lower than the dimension of the observed variables; i.e. $p \gg q$. The objective is thus to find the principal components, which are a linear combinations forming what is known as the loading matrix denoted by $A \in \mathbb{R}^{p \times q}$. This is achieved by searching for the directions that result in PC scores with maximum variances. In doing so, the resulting principal components and their variances correspond to the eigenvectors and eigenvalues of the covariance matrix $\Sigma$ of the observations $x$.

Traditional PCA deals with the $L_2$ norm, which is optimal when we are concerned with minimizing the Mean Square Error (MSE) (Wang et al. 1996). However, when PCA is applied in noisier environments with significant outliers, it becomes increasingly important for a more robust measure. In some cases, outliers can be removed from the estimation process, making regular PCA adequate. However, in practice we do not know which data points are outliers, especially in a high dimensional setting. Not to mention that in such a setting, each data point is too valuable to be discarded. This is due to the usual scarcity of observations relative to the dimension. In the literature, the problem of modeling data sets with erroneous entries and outliers while simultaneously detecting them is referred to as the robust PCA problem.

Iterative Reweighted Least Squares (IRLS) is a straight forward algorithm for obtaining robust components (De La Torre and Black 2003). The basic idea is to iteratively apply regular PCA while down-weighting poorly fitted observations between iterations. Another approach proposed by Candès et al. (2011) assumes that the data is a superposition of a low rank and a sparse component. The objective is then to find the decomposition that minimizes a weighted mixture of the nuclear norm and the $L_1$ norm. This method solves a convex program called the Principal Component Pursuit (PCP). A detailed review of these algorithms as well as other variations and extensions can be found in (Vidal et al. 2016).

Classical and robust variations of PCA obtain a lower dimensional subspace that is spanned by a linear combination of all the variables in the original high dimensional subspace. This results in interpretability issues especially in data rich environments. This shortcoming of PCA can be addressed by adjusting the original formulation such that a relatively small number of nonzero entries are allowed for the loadings. These nonzero elements will thus correspond to the features that contribute the most information in the data population. Such formulations are commonly referred to as Sparse Principal Component Analysis (SPCA) (Zou et al. 2006).

Several methods for obtaining these sparse principal components have been explored in the literature. One intuitive approach that was proposed by Cadima and Jolliffe (1995), is to threshold loadings with a small absolute value to zero. This is generally referred to as "simple thresholding" (d'Aspremont et al. 2008). The Simplified Component Technique-LASSO (SCoTLASS) introduces a bound on the sum of the loadings, thereby forcing some of them to become zero (Jolliffe et al. 2003). The sparse low-rank approximation (SLRA) algorithm proposed in (Zhang et al. 2002, 2004) computes matrix low-rank approximations with sparse factors, which is then formulated as a penalized optimization problem. Another SPCA algorithm, introduced by Zou et al. (2006), reformulates the traditional PCA problem as a regression problem. Then, it adds a LASSO (Tibshirani 1996) type penalty, which is a penalized regression technique based on the $L_1$ norm. Both previous formulations result in non-convex optimization problems that can cause computational issues. d'Aspremont et al. (2008) later introduced another approach that directly incorporates a sparsity condition in the SPCA problem formulation, which resulted in a convex relaxation of the original problem. More recently, Ma (2013) proposed a new iterative thresholding approach. Under a spiked covariance model, this approach was shown to obtain the leading principal components more consistently in sparse high-dimensional settings.

High dimensional data often contain outliers as well as sparse data structures. Few work has been done to combine the two properties of robust and sparse principal component analysis. Most notably Robust Sparse Principal Component Analysis (RSPCA) and (ROSPCA) introduced in (Hubert et al. 2016, Croux et al. 2013). The former combines the two properties by applying the $L_1$ penalty to the projection pursuit (PP) approach for obtaining robust principal components. While the latter incorporates sparse PCA within the framework of the robust method ROBPCA, by first finding a robust subspace and then uses the sparse method SCoTLASS to yield a sparse loading matrix. These two approaches for finding both sparse and robust subspaces will serve as a benchmark in the simulation study of section 4.

## 2.2 Probabilistic PCA

The classical formulation of principal component analysis is not a probabilistic model in the sense that the latent variable $z$ is viewed as a projection of the observed variables $x$ onto a linear correlation subspace of interest. Therefore, it emphasizes more the observed variables $x$ rather than the latent variables $z$. Probabilistic principal component analysis uses a probabilistic generative model that was introduced by Tipping and Bishop (1999b). It is called a probabilistic generative model because it models the observed variables as if they are generated from the latent variables with some Gaussian error, while assigning probability distributions to them. This puts the latent variables $z$ in the forefront while the observed variables $x$ are treated as a byproduct that results from a linear combination of the latent variables $z$.

Based on this probabilistic interpretation of PCA, other variations using Bayesian variational inference to improve robustness have been proposed in the literature (Archambeau et al. 2006, Gao 2008). A latent variable view of the Student-t distribution is used by Archambeau and Bach (2009) instead of choosing a Gaussian noise for the error. While, Gao (2008) replaces the conventional Gaussian distribution for the noise by the Laplacian distribution, also referred to as the $L_1$ distribution. Both distributions are characterized by their heavy tails which allows them to be more robust to outliers as opposed to the traditional Gaussian distribution. The use of these alternative priors will be detailed further in the methodology part of this article in section 2.3.

In addition to the formulations mentioned above, a different probabilistic implementation that focuses on decomposing the latent variable into a low-rank component and a sparse component can be found in (Han et al. 2017, Ding et al. 2011). While both studies propose methods for solving this decomposition, Han et al. (2017) assumes that the sparse component is structured rather than consisting of independent variables.

Variants of PCA probabilistic formulations, that promote sparsity, have also been discussed in the literature. Guan and Dy (2009) proposed using three different sparsity inducing priors (Laplacian, inverse-Gaussian and Jeffrey's prior) for the loading matrix in a Bayesian probabilistic formulation of PCA. In a more general framework, a probabilistic projection model was introduced in (Archambeau and Bach 2009), with an application to sparse PCA as a special case. The use of these alternative priors will be detailed further in the methodology part of this article in section 2.3.

## 2.3 Bayesian Variational Inference for RSPCA

The objective of Bayesian variational inference is to approximate the posterior distributions of the hidden variables $p(\Theta|x)$. Let $g(\Theta)$ represent this approximation, also referred to as the variational distribution (Gao 2008). The KL-divergence between $g(\Theta)$ and $p(\Theta|x)$ is the difference between the log marginal likelihood $p(x)$ and a lower bound given by:

$$\mathcal{L}\big(g(\Theta)\big) = \int g(\Theta)\ln\frac{p(x,\Theta)}{g(\Theta)}d\Theta \leq \ln \int p(x,\Theta)d\Theta = \ln p(x). \qquad (1)$$

Thus minimizing the KL divergence is equivalent to maximizing this lower bound, also known as the negative free energy in statistical physics (Gao 2008). The objective then becomes to minimize the Kullback–Leibler (KL) divergence between $g(\Theta)$ and true posterior distribution $p(\Theta|x)$ as follows:

$$\max_{g(\Theta)} \mathcal{L}\big(g(\Theta)\big) \equiv \min KL(g(\Theta)||p(\Theta)) = \int g(\Theta)\ln\frac{g(\Theta)}{p(\Theta|x)}d\Theta. \qquad (2)$$

The next step is to postulate a simple parameterized family of distributions over $g(\Theta)$ such that it is tractable to evaluate the negative free energy while simultaneously obtaining a tight lower bound. The mean field family of distributions is one where $g(\Theta)$ factorizes over all the parameters as follows:

$$g(\Theta) = g(\theta_1)g(\theta_2)g(\theta_3)\ldots \qquad (3)$$

The Bayesian variational inference procedure consists of two steps that are akin to the Expectation-Maximization (EM) algorithms. In the first step (E-step), the variational parameters are fixed at Θ, while the variational distribution $g(\theta)$ is updated to minimize the KL-divergence. The resulting variational posteriors as such:

$$g(\theta) \propto e^{E_{\Theta|\theta}[\ln p(X, \Theta|\theta)]}. \qquad (4)$$

Here, $g(\theta)$ is the posterior of $\theta \in \Theta$ and $E_{\Theta|\theta}$ denotes the expectation with respect to all the parameters in the set Θ excluding the parameter $\theta$, which is fixed while its respective posterior is calculated.

For the second step (M-step), the updated variational posterior distribution obtained from the first step is fixed. Then the variational set of parameters Θ are updated by maximizing the lower bound given the factorized variational distributions. The two steps are iterated, and the parameters are updated sequentially while the remaining are fixed.

## 3. RS-PCA METHODOLOGY

The following is a description of the problem we address in this article. Suppose we are observing a process with $p$ variables. The observed data at time $t$ is represented by the $p$ dimensional vector $\boldsymbol{x_t} = (x_{1,t}, \dots, x_{p,t})'$. Our objective is to monitor the data acquired and detect any shifts in any components of the data streams. In our analysis we assume that the data can be projected on a lower dimensional subspace spanned by a set of latent variables $\boldsymbol{z_t} = (z_{1,t}, \dots, z_{q,t})'$. Then the objective of the study is twofold: (1) to find the projection matrix (principal components) from in-control historical data, (2) to use the projections of online data streams onto the latent space for process monitoring.

The following subsections address the aforementioned objectives. Subsections 3.1 and 3.2 discuss how to extract the feature space by formulating and probabilistically solving a robust and sparse representation of PCA. While, subsections 0 and 3.4 propose the monitoring and diagnostic schemes based on the extracted features.

## 3.1 Probabilistic Robust and Sparse Model Formulation

In this paper we assume that the observed data $x \in \mathbb{R}^p$ and the latent (feature) variables $z \in \mathbb{R}^q$ satisfy the following model:

$$x = A z + \varepsilon. \tag{5}$$

Here, $A \in \mathbb{R}^{p \times q}$ is the loading matrix and $\varepsilon$ is the noise. We aim to find a sparse representation of the loading matrix $A$ from the observed data, so that we can obtain a sparse representation of the data. For that purpose, we assume that the covariance matrix $\Sigma$ of the data streams is block diagonal. In other words, data streams fall into $K$ disjoint sets $B = \{1, \dots, K\}$ such that $\text{cov}(x_{i,t}, x_{j,t}) = 0$ if $X_{i,t}$ and $X_{j,t}$ belong to two different sets. This assumption is directly related to the interpretability of the feature space. Thereby, each estimated principal component can only be a linear combination of variables belonging to a single set. Hence, the sparsity of the components dictates the interpretability of the model.

To probabilistically induce sparsity to the model given by equation (5), we impose a Laplacian prior onto the elements of loading matrix $A$ as discussed in section 2.2. Mathematically the prior is as follows:

$$p(A_{i,j}|\lambda) = \sqrt{\frac{1}{2\lambda}} \exp\left\{-\sqrt{\frac{2}{\lambda}}|A_{i,j}|\right\}, \qquad i \in \{1, \dots, p\}, j \in \{1, \dots, q\}. \tag{6}$$

Where, $A_{ij}$ is the $(i,j)$ element of the loading matrix $A$ and the parameter $\left(\sqrt{\frac{\lambda}{2}} > 0\right)$ is the scale parameter of the Laplacian distribution.

Moreover, to provide robustness to equation (5), we propose a Laplacian prior for the noise $\varepsilon$ as follows:

$$p(\varepsilon_i | \gamma^{-1} = \sigma_\varepsilon^2) = \sqrt{\frac{\gamma}{2}} \exp\{-\sqrt{2\gamma}|\varepsilon_i|\}, \quad i \in \{1, \dots, p\}. \tag{7}$$

Here, the parameter $\left(\sqrt{2\gamma} > 0\right)$ is the scale parameters of the Laplacian distribution. The priors given by equations (6) and (7) offer the combined robust and sparse properties.

Note that the novelty of our approach is in introducing the Laplacian priors given by equations (6) and (7) to the loading matrix $A$ and the noise $\varepsilon$ to promote robustness and induce sparsity simultaneously. Traditionally, the assumption for the noise of the observed data is the Gaussian distribution, as described in subsection 2.2. The popularity of the Gaussian prior can be attributed to the attractiveness of the central limit theorem, and that it facilitates solving the model with a simple expectation maximization algorithm. However, it may be inappropriate in practice when the data is contaminated with outliers. Meanwhile, the Laplacian prior offers robustness against outliers, but it is not a conjugate to the Gaussian distribution, and therefore poses new computational challenges for computing the posteriors.

Defining the priors for the loadings $A_{i,j}$ and noise $\varepsilon_i$ as presented by the equations above may lead to an intractable formulation. Fortunately, the Laplacian distribution can be represented as an infinite superposition of Gaussian distributions (Archambeau and Bach 2009, Gao 2008). This inspired us to introduce intermediate variables $\Lambda_{i,j}$ and $\Gamma$ in the following manner:

$$p(A_{i,j}|\lambda) = \int p(A_{i,j}|\Lambda_{i,j}^{-1})p(\Lambda_{i,j}|\lambda)d\Lambda_{i,j}$$

$$= \int \sqrt{\frac{\Lambda_{i,j}}{2\pi}} exp\left\{-\frac{\Lambda_{i,j}A_{i,j}^2}{2}\right\} \frac{1}{\lambda\Lambda_{i,j}^2} exp\left\{-\frac{1}{\lambda\Lambda_{i,j}}\right\} d\Lambda_{i,j}$$

$i \in \{1,\ldots,p\}, j \in \{1,\ldots,q\}.$  (8)

$$p(\varepsilon_i|\gamma^{-1}) = \int \sqrt{\frac{\Gamma\gamma}{\pi}} exp\{-\Gamma\gamma\varepsilon_i^2\} p(\Gamma)d\Gamma$$

$i \in \{1,\ldots,p\}.$  (9)

and $p(\Gamma) = \frac{1}{2\Gamma^2} exp\left\{-\frac{1}{2\Gamma}\right\}$

In equations (8) and (9) we decompose the Laplacian distribution into a two-level hierarchy. The first level is to impose Gaussian distribution priors on $A_{i,j}|\Lambda_{i,j}^{-1}$ and $\varepsilon_i|\Gamma^{-1}$. The second level is to impose inverse Gamma distributed hyper-priors on the intermediate variables $\Lambda_{i,j}$ and $\Gamma$ with their respective scale parameters $\frac{2}{\lambda}$ and $\frac{1}{2\gamma} > 0$, while the shape parameter is set to 1. To visualize how the variables in this reformulated model relate, Figure 2 shows a complete graphical representation.

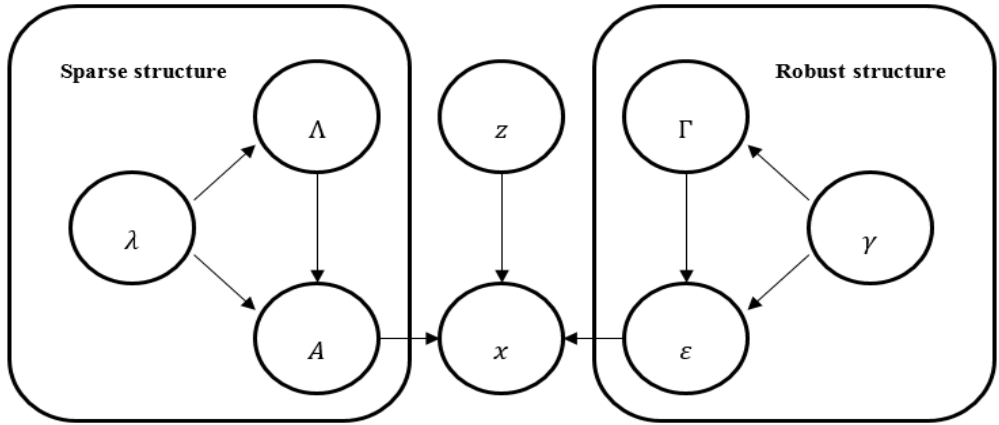

Figure 2 RSPCA graphical model (notation details in text). Arrows indicate conditional dependencies between model variables and parameters

It is worth noting that if we marginalize out the first level priors $p(\Lambda_{i,j})$ and $p(\Gamma)$ in equations (8) and (9) we retrieve equations (6) and (7), respectively. The prior for the latent variable $z$ is left as it is in the probabilistic PCA model as $z \in \mathbb{R}^q \sim \mathcal{N}(0, \Phi^{-1})$, where $\Phi$ is a diagonal covariance matrix. Finally we impose a $Gamma(a, b)$ prior on $\gamma = \sigma_\varepsilon^{-2}$.

Given the alternative priors presented in equations (8) and (9), the joint distribution of the observation set $x$, latent variable, loading matrix, hidden parameters and hyperparameters $\{\Theta \equiv z, A, \Lambda, \lambda, \Gamma, \gamma\}$ becomes:

$$p(x, \Theta) = \prod_{t=1}^{T} p(x_t - Az_t | A, \Theta) p(z_i)$$
$$\times \prod_{i=1}^{p} p(\varepsilon_i | \Gamma^{-1}) p(\Gamma | \gamma) p(\gamma) \prod_{j=1}^{q} p(A_{i,j} | \Lambda_{i,j}^{-1}) p(\Lambda_{i,j}). \quad (10)$$

We are interested in evaluating the posterior distributions of the hidden variables given the observations. However, the posteriors are computationally intractable because the marginal distribution $p(x)$ cannot be obtained analytically. The next subsection discusses how to circumvent this issue.

### 3.2 Variational Posteriors and Update Equations

In this subsection, we utilize Bayesian variational inference as an approximation tool for estimating the posteriors. The chosen variational distribution for the latent variable $g(z_t)$ is the Gaussian density. The expectation over the variational posterior of a parameter $\theta$ will be represented by $\langle \theta \rangle$ for notation purposes. The variational posterior density will also be Gaussian with the following update equations for the mean $\langle z_t \rangle$ and variance $\Sigma_z$, respectively:

$$\langle z_t \rangle = \langle \gamma \rangle \langle \Gamma_t \rangle \Sigma_z \langle A \rangle^T x_t, \quad (11)$$

$$\Sigma_{z_t} = (\Phi + \gamma \langle A^T \Gamma_t A \rangle)^{-1}, \tag{12}$$

Next, we look at the loading matrix $A$, where every row $A_{i.}$ is set to have independent Gaussian variational distribution. Hence the joint distribution of the rows of $A$ is as follows:

$$g(A) = \prod_{i=1}^{p} g(A_{i.}), A_{i.} \in \mathbb{R}^q \sim \mathcal{N}(\langle A_{i.} \rangle, \Sigma_{A_{i.}}). \tag{13}$$

The respective update equations for the mean $\langle A_{i.} \rangle$ and variance $\Sigma_{A_{i..}}$ of the resulting Gaussian posterior are given by:

$$\langle A_{i.} \rangle = \langle \gamma \rangle \Sigma_{A_{i.}} \sum_{t=1}^{n} \langle z_t \rangle^T x_{.i}, \tag{14}$$

$$\Sigma_{A_{i.}} = [\text{diag}(\langle \Lambda_{i.} \rangle) + \gamma \sum_{t=1}^{n} \langle z_t^T \Gamma_t z_t \rangle]^{-1}. \tag{15}$$

As discussed in (Archambeau and Bach 2009, Gao 2008), the variational distribution for the hidden parameters $\Lambda_{ij}$ is set as the generalized inverse Gaussian distribution given by:

$$g(\Lambda_{ij}) \sim GIG(\omega, \chi, \psi) = \frac{\chi^\omega (\sqrt{\chi \psi})^\omega}{2 K_\omega(\sqrt{\chi \psi})} \Lambda_{ij}^{\omega-1} exp\left\{-\frac{1}{2}\left(\chi \Lambda_{ij}^{-1} + \psi \Lambda_{ij}\right)\right\}. \tag{16}$$

Here, $\omega = -\frac{1}{2}$ is the index, $\chi = l_{ij}$ and $\psi = \frac{2}{\lambda}$ are shape parameters, where $l_{ij}$ is the $j^{\text{th}}$ diagonal element of $\langle A_{i.}^T A_{i.} \rangle$. Moreover, $K_\omega(\cdot)$ represents the modified Bessel function of the second kind. The choice of the index of and shape parameters given in equation (19) reflects the priors on $p(\Lambda_{ij})$ and $p(A_{ij}|\lambda)$ described in subsection 3.1. Then, the update of $\Lambda_{ij}$ is:

$$\langle \Lambda_{ij} \rangle = \sqrt{\frac{\chi}{\psi}} \frac{K_{\omega+1}(\sqrt{\chi \psi})}{K_\omega(\sqrt{\chi \psi})} = \sqrt{\frac{\lambda l_{ij}}{2}}. \tag{17}$$

Subsequently, we discuss the variational distribution of the hidden intermediate parameter of the noise $\Gamma_t$. The choice for it is also the generalized inverse Gaussian distribution $GIG\left(-\frac{1}{2}, 1, \gamma\eta_t\right)$, as is the case for $\Lambda_{ij}$. This is because they both have the same two-level hierarchical structure, described in equations (7) and (8). Here, $\eta_t$ is:

$$\eta_t = \frac{1}{p}\text{tr}\left[(x_t - \langle Az_t\rangle)(x_t - \langle Az_t\rangle)^T + A\Sigma_{z_t}A^T\right], \tag{18}$$

where, tr(·) refers to the trace of the enclosed matrix. Furthermore, the updates for $\Gamma$ has the following form:

$$\langle \Gamma_t \rangle = \sqrt{\frac{1}{\gamma\eta_t}}. \tag{19}$$

Finally, the variational distribution for the reciprocal of the error noise $\gamma$ is still a $Gamma(a, b)$ with mean $\langle \gamma \rangle = \frac{a}{b}$. The update equations for the hyperparameters $\gamma$ and $\lambda$, respectively are:

$$a \leftarrow a + \frac{np}{2}$$
$$b \leftarrow b + \frac{1}{2}\sum_{t=1}^{n}\left[(x_t - \langle Az_t\rangle)\langle\Gamma_t\rangle(x_t - \langle Az_t\rangle)^T + \text{tr}(\langle\Gamma_t\rangle A\Sigma_{z_t}A^T)\right], \tag{20}$$

$$\lambda \leftarrow \frac{1}{pq}\sum_{i=1}^{p}\sum_{j=1}^{q}\Lambda_{ij}. \tag{21}$$

| Algorithm 1: Probabilistic (RS-PCA) Via Bayesian Variational Inference | |
|---|---|
| | Input: $\lambda, \gamma(a, b), A$ |
| | From historical data $\boldsymbol{x_t}, t = 1, \ldots, n$ |
| 1 | Calculate the posterior distribution of $\boldsymbol{z_t}$ according to eq. (11) and (12) given the current estimate of $A$. |
| 2 | Update the elements of $\Lambda$ using eq. (17). |
| 3 | Update the parameter $\lambda$ with eq. (21). |
| 3 | Calculate the posterior of $A$ according to eq. (14) and (15). |
| 4 | Update the parameter $\gamma(a, b)$ with eq. (20), and $\Gamma$ using eq. (19). |
| | Repeat steps 1-4 until convergence is achieved. |

In summary, the above Algorithm 1 gives an overview of the Bayesian variational inference procedure described in this subsection. The initial parameterization of the model can be obtained using traditional probabilistic PCA.

## 3.3 RSPCA based Process Monitoring

The critical step in process monitoring is the evaluation of probability densities of all the variables. The probabilistic formulation of RSPCA that was presented in the previous subsections lays down the foundation for process monitoring. The probabilistic approach used in this article to model RSPCA assumes the densities of $p(z), p(\varepsilon)$ and $p(A)$ as discussed in subsection 3.1. What remains is to evaluate $p(x)$, as well as the posteriors $p(z|x)$ and $p(\varepsilon|x)$. For process monitoring using regular probabilistic PCA (Kim and Lee 2003), these densities are directly derived from using the probabilistic formulation of PCA given in equation (5). However, a similar direct approach is not possible for our proposed formulation as the posteriors cannot be obtained in a straightforward fashion. Therefore, we utilize the variational densities and their posteriors, which were discussed in subsection 3.2, to approximate the true densities.

### 3.3.1 Monitoring Latent Variables

We begin by discussing monitoring the latent variables $z$, which depending on the application may represent a fault pattern or a specific physical phenomenon that associates multiple observation variables $x$. Our assumption for the distribution of the latent variables in the probabilistic model is that $z \sim \mathcal{N}(0, \mathbf{\Phi}^{-1})$. Since we cannot observe the latent variables directly, we propose to estimate them using their variational posterior densities $g(z)$. Given a new sample $x_\tau$, the hypothesis test for whether the sample is in-control is as follows:

$$H_0: z_\tau = 0$$
$$H_1: z_{\tau,i} \neq 0, \quad \text{For at least one } i \text{ such that } i \in \{1, \ldots, q\}. \tag{22}$$

Since we are dealing with Gaussian density for $z_{\tau,i}$, the test statistic is $X_0 = \langle z_t \rangle^T \Sigma_z^{-1} \langle z_t \rangle$ and we propose to reject the null hypothesis $H_0$ given in (22) if $X_0 = \left\| \langle \gamma \rangle \langle \Gamma_t \rangle \Sigma_z^{1/2} \langle A \rangle^T x_t \right\|^2 > \chi^2_{(1-\alpha,q)}$.

### 3.3.2 Monitoring the noise variable

Monitoring the latent variable is useful for detecting out of control instances, which is the case only when the instance is in accordance with the RSPCA model developed in section 3.1. This is where monitoring the noise variable comes into play. It identifies instances that do not share the same subspace structure. This is similar to the Q-statistic that complements the hoteling-$T^2$ statistic. In a similar fashion to monitoring the latent variable, the distance between the incoming sample and the model will be estimated using the mean of the variational posterior density and the hypothesis test then becomes:

$$H_0: \varepsilon_\tau = 0$$
$$H_1: \varepsilon_{\tau,i} \neq 0, \quad \text{For at least one } i \text{ such that } i \in \{1, \ldots, q\}. \tag{23}$$

The test statistic is then $X_0 = \gamma \langle \varepsilon_\tau \rangle^T \langle \varepsilon_\tau \rangle$ and we propose to reject the null hypothesis given by (23) if $X_0 = \left\| \langle \gamma \rangle^{-1/2} \langle \Gamma_t \rangle^{-1/2} (x_t - \langle A z_t \rangle) \right\|^2 > \chi^2_{(1-\alpha,p)}$.

### 3.4 Fault Diagnosis

Once an incoming sample has been identified to be an out-of-control instance, it is desired to isolate the responsible variables for this irregularity. This diagnostic step can be quite difficult since our detection of the out-of-control instance is based on the hidden latent variables rather than the observed variables themselves. Therefore, the diagnosis method should be able to distinguish the observable variables that contribute towards the irregularity detected in the latent variable subspace.

The diagnostic procedure can be decomposed into the following steps: (1) identifying the out-of-control latent variable $(z_j, j \in \{1, ..., q\})$ and (2) determining the set $S \in \mathbb{R}^s$ of observable variables that contribute to the identified latent variable $z_j$. It is important to mention that for the remainder of this study we assume that only one latent variable can go out-of-control at any given time in the steps mentioned previously. This could be interpreted as each fault type being associated with a single latent direction.

Isolating the out of control latent variable $z_j$ can be achieved via decomposition methods for the Hotelling $T^2$ statistic of the hypothesis test (22). The most common method would be the Mason-Tracey-Young (MTY) method that was proposed in (Mason et al. 1995). This decomposition relaxes to finding the latent variable $z_j$ that has a significant deviation from the mean. In other words, $z_j$ such that $(\langle \Gamma \rangle \Sigma_z \langle A \rangle^T x_\tau)_j^2 > \frac{\tau+1}{\tau} F_{1,\tau-1}$.

Determining the set of observable variables that contribute to the out-of-control latent variable $z_j$ is not as straightforward. This is because $A_{ij}$ is a random variable, and finding the set of $x_i \in S$ such that $A_{ij} \neq 0$ corresponds to the following hypothesis test:

$$H_0: A_{ij} = 0$$
$$H_1: A_{ij} \neq 0 \tag{24}$$

The hypothesis in (24) is rejected when the test statistic $\left[\left(\Sigma_{A_{i.}}\right)_{jj}\right]^{-1/2} \left[\langle\gamma\rangle\Sigma_{A_{i.}} \sum_{t=1}^{n}\langle z_t\rangle^T x_{.i}\right]_j \leq t_{\tau-1}$. Here, $\left(\Sigma_{A_{i.}}\right)_{jj}$ is the $j$th diagonal element of $\Sigma_{A_{i.}}$, which corresponds to the marginal variance of $A_{ij}$. For all $A_{ij}$ ($i = 1, \ldots, p$) such that the hypothesis is rejected, we conclude that the observed variable $x_i \in S$ contributes to the detected out-of-control $z_j$ instance.

## 4. SIMULATIONS

This section presents the results of simulation studies to validate our proposed method and to test its monitoring performance. The first subsection 4.1 discusses the method in which we generate the simulation data. The following subsection 4.2 illustrates the ability of the proposed methodology to accurately recover the loading matrix and compares it to the state-of-the-art methods, which also serves as a verification step. The final subsection 4.3 evaluates the monitoring capability using the proposed methodology, while comparing it to other benchmark dimension reduction techniques.

The robust and sparse properties of our proposed methodology are evaluated in this section against state of the art techniques from the literature. Two benchmark methods are considered in the simulated experiments. The first method by Croux et al. (2013) will be denoted as SRPCA, and the second being ROSPCA (Hubert et al. 2016). Initially, we describe the simulated data generation procedure in subsection 4.1.

### 4.1 Data Generation

The model given by (5) is the base of the data generation method used in the following simulations. We adopt a setup consistent with the one described in (Hubert et al. 2016) for generating the data. First a constant loading matrix $A$ is generated. To promote sparsity in the covariance of the observation variables $x$, the columns of $A$ are designed to be sparse in a block-wise fashion with $K$ blocks. Block $B_k$ has cardinality $|B_k| = b_k$ such that the corresponding loading $A_{i,j}$ for all $x_i$ and $z_j$ becomes:

$$A_{i,j} = \begin{cases} -\frac{1}{\sqrt{b_k}} & \text{if } x_i \text{ and } z_j \in B_k \text{ for some } k \\ 0, & \text{otherwise} \end{cases} \quad (25)$$

The nodes represent the observable and latent variables, while edges represent a non-zero element of the loading matrix. The cardinality of the first two blocks are chosen to be the same (i.e. $b_1 = b_2 = b$), while the remaining blocks are unit blocks ($b_k = 1$ for all $k > 2$). The total number of blocks, which is also the number of latent variables, is thus $K = p - 2b + 2$.

Figure 3 provides a network visualization of the blocks. Next, the latent variables $z_j$ are generated from a normal distribution with zero mean and a diagonal covariance matrix $\Sigma_z$. The variances of $z_1$ and $z_2$ are set to be significantly larger than the remaining latent variables so the principal components corresponding to them can be identified as the first and second, respectively. Finally, white noise is added to normal observation and $100\delta\%$ of the observations are replaced by outlier observations. Outliers are independently generated from a multivariate normal distribution with mean $\mu_{out}$ and diagonal covariance matrix $\sigma_{out}^2 I_p$. The outliers are generated such that they do not follow the correlation structure of the normal observations which will emphasize the need for robustness.

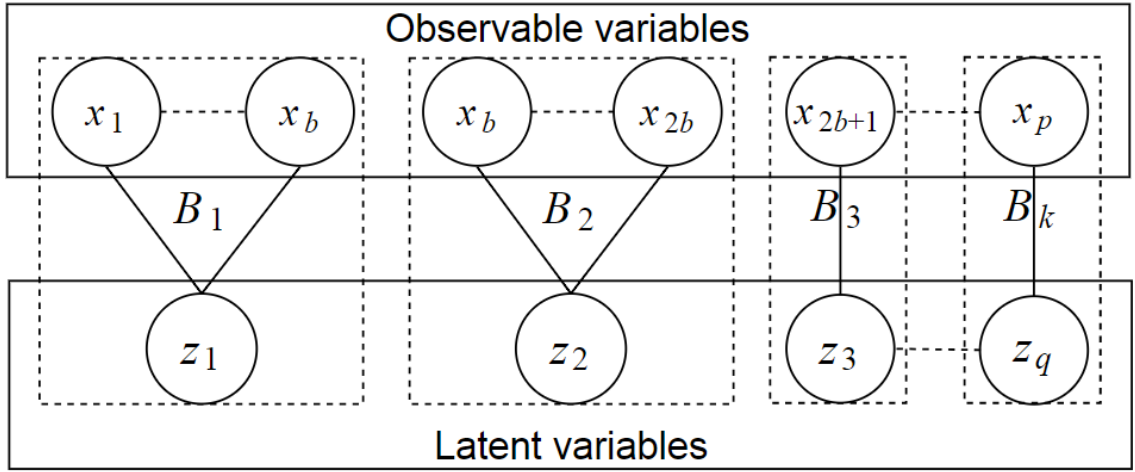

Figure 3 Network visualization of the simulation data generation blocks

## 4.2 Loading Matrix Recovery Experiment

In this subsection, we evaluate the recovery of the loading matrix by generating a set of data as described in the previous subsection. We consider a high-dimensional setting with $p = 500$ and $n = 50, 100, 500$. Only cases $n \leq p$ where considered as the difference in the accuracy of retrieving the principal components via the different methods becomes negligible. Moreover, having the sample size be smaller than the dimension of the data is the challenging scenario of interest in this study. This is consistent with benchmark studies and it is also similar to the motivating Buckypaper manufacturing process monitoring problem, where the dimension of Raman spectra is 512. The first two blocks have size $b = 20$ and the variances of the latent variables $\text{diag}(\Sigma_z) = [233, 49, 4(422 \text{ times}), 2(19 \text{ times}), 0.4(19 \text{ times})]$. Outliers with mean $\mu_{out} = 25(0, -4, 4, 2, 0, 4, -4, 2, \dots \text{ (for the first two blocks)}, 3, -3, \dots, 3, -3)^T$ , variance $\sigma^2_{out} = 20$ and proportions $\delta = 0.1, 0.2, 0.3$.

We simulate 100 datasets for each experimental setting to keep computations practical. To compare the robustness of our proposed RSPCA method against SRPCA and ROSPCA, we utilize the average deviation angle as was used in the benchmark studies. This computes a measure between the estimated subspace and the true subspace and results in an angle between 0 and $\frac{\pi}{2}$, which is then normalized to produce a measure between 0 and 1. Values closer to 0 are desired as they represent a closer estimate. The tuning parameter that controls sparsity for SRPCA and ROSPCA is chosen based on the BIC criterion proposed in the respective study. For ROSPCA, the parameter determining the degree of robustness, which constitutes a lower bound on the number of normal observations, is set to 0.5 for maximal robustness as suggested in (Hubert et al. 2016). The average deviation angle (standard deviations) results from the experiments are summarized in Table 1.

Table 1. Average deviation angle (standard deviation) values of extracted PCs

| $\delta$ | 0.1 | | | 0.2 | | | 0.3 | | |
|---|---|---|---|---|---|---|---|---|---|
| $n$ | 50 | 100 | 500 | 50 | 100 | 500 | 50 | 100 | 500 |
| RSPCA | 0.61 (0.09) | 0.36 (0.06) | 0.15 (0.03) | *0.63* (0.12) | *0.40* (0.09) | *0.16* (0.05) | *0.69* (0.14) | *0.45* (0.15) | 0.18 (0.08) |
| ROSPCA | *0.59* (0.11) | *0.34* (0.08) | *0.14* (0.04) | 0.64 (0.14) | 0.42 (0.10) | 0.17 (0.05) | 0.70 (0.17) | 0.46 (0.14) | *0.18* (0.07) |
| SRPCA | 0.71 (0.18) | 0.44 (0.13) | 0.15 (0.06) | 0.75 (0.21) | 0.51 (0.15) | 0.18 (0.08) | 0.82 (0.32) | 0.58 (0.13) | 0.19 (0.17) |

The simulation results indicate that the estimation of the principal components improves as $n$ increases in terms of both bias and variance. Our proposed probabilistic approach appears to yield better results than SRPCA, while being competitive with ROSPCA and even slightly outperforming it in the case of moderate outliers ($\delta = 0.2$).

It remains to evaluate the capability of the different methods in correctly identifying the sparse structure of the principal directions. We use the *zero measure* to compare those three techniques.

The *total zero measure* is the proportion of loadings correctly identified as 0 or nonzero. For SRPCA and ROSPCA, an element of a loading matrix is considered to be 0 if its absolute value is smaller than $10^{-5}$. While for our proposed RSPCA method, we test the hypothesis in (24) at significance level 0.01 to determine whether an element is 0 or not.

Figure 4 illustrates the total zero measure for RSPCA against the benchmark methods. It can be seen that our proposed probabilistic approach is superior in distinguishing the sparse structure of the principal subspace even in the case of high contamination levels and low sample sizes. This is because the probabilistic model allows for a better way to discern zero loadings via hypothesis testing. This takes into account the variances in the estimates rather than uniformly thresholding all values.

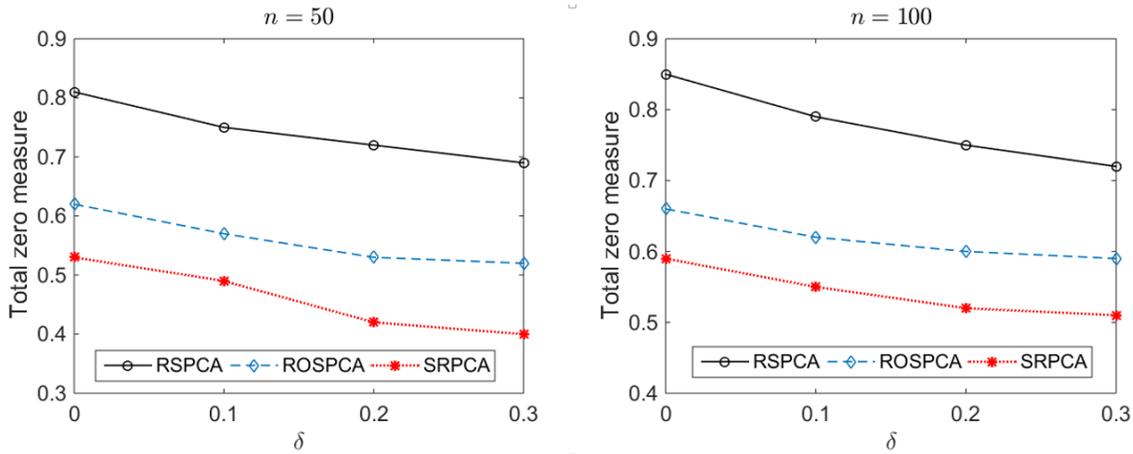

Figure 4 Total zero measure of proposed RSPCA vs. benchmark methods ROSPCA and SRPCA

## 4.3  Monitoring Performance

This subsection is aimed towards testing the performance of our method in detecting changes that occur while monitoring a process. To better demonstrate the need for a monitoring procedure with both robust and sparse properties, we compare the detection delay to that of classical PCA, sparse PCA and robust PCA. The implementation of the other variations is based on modifications to our own method to remove robustness, sparsity, or both during the extraction of principal components. This self-implementation is similar to the proposed approaches in the literature (Zeng et al. 2017, Ge and Song 2011, Kim and Lee 2003). In control data sets are generated using the same method from the previous subsections, while out-of-control observations are generated by shifting the first latent variable $z_1$ by a range from 0 to 2 standard deviations from the mean. The principal components are learned from $n = 500$ in control observations with outliers with contamination proportion $\delta = 0.1, 0.2$. The average detection delays from 100 iterations using the different approaches are summarized in Figure 5, where the in-control average run length is set to 200.

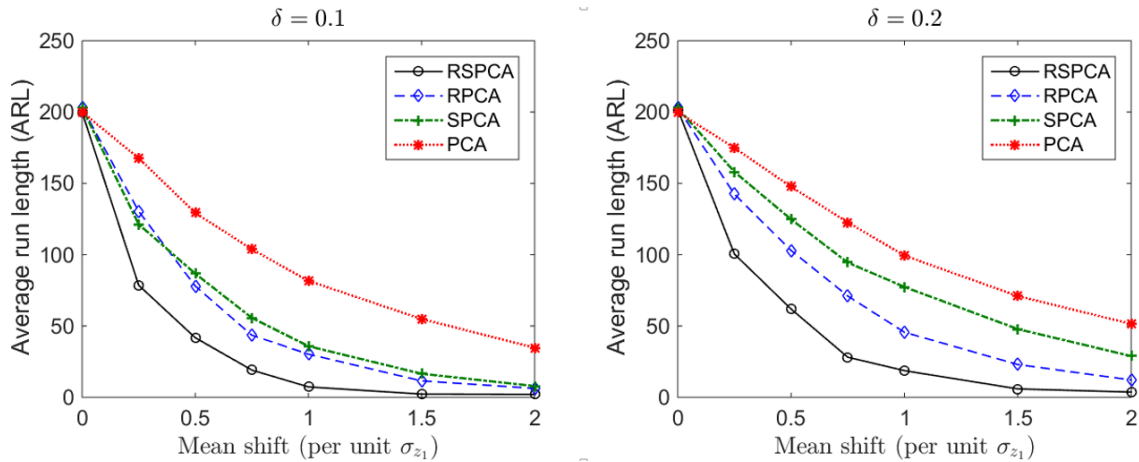

Figure 5 Illustration of the detection delay

From the results of Figure 5, it can be seen how the detection delay has been decreased significantly when using our proposed approach, which considers both robustness and sparsity to improve the change detection monitoring capability. At the lower contamination level $\delta = 0.1$ the performance of RPCA and SPCA is relatively similar especially at very small mean shifts.

However, the advantages of robust estimation becomes more clear at the higher contamination level $\delta = 0.2$, which is to be expected. Classical PCA performs relatively poorly in all settings since it does not take into account the sparse structure or the outliers in the training data.

## 5. CASE STUDY

In this subsection we test our proposed methodology in addressing the challenges of monitoring the production process of continuous CNTs buckypaper using inline Raman spectroscopy. We aim to show that the sparsity and robust properties of our proposed method can address the sparse peak locations and the complex noise structure.

The data we use in our case study is from a surrogated Raman spectra from a practical experiment. More details on the experiment and the data acquisition can be found in (Yue et al. 2018). Figure 6 illustrates the Raman spectra obtained from the experiments. The highlighted regions correspond to the sparse segments where defects occur. The first highlighted segment represents the D-band while the third is the G-band. These bands contain relevant quality information such as molecular defects in the CNT structure as well as functionalization (Cheng et al. 2010). Therefore, it is important to be able to detect irregularities in these segments when monitoring the whole profile. Defects, which are not associated with either the D-band or G-band, can also occur in other regions such as the middle region highlighted in the figure. From the zoomed in box of profiles, we can note the existence of outlier observations (solid blue profiles) with excessive noise that appear to mask the defects.

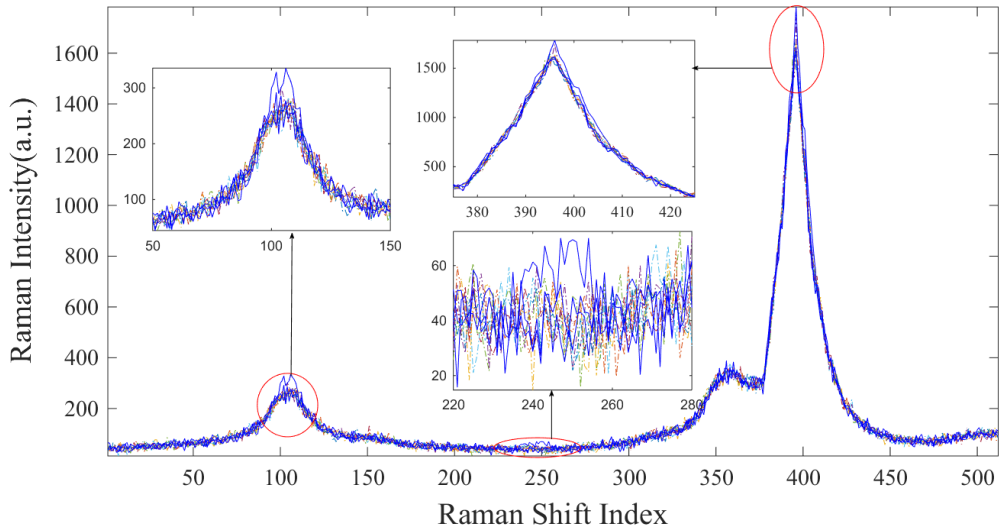

Figure 6 Illustration of the Raman spectra data

We begin the monitoring procedure by extracting the principal components. The components that most explain the variance in each of the sparse segments are respectively shown in Figure 7 for the different PCA methods. We can see that our proposed robust and sparse method successfully isolates the sparse segments similar to sparse PCA. While the ones obtained by regular PCA and robust PCA are mixed with other regions. Moreover, the inherent robustness property of our method provides a better representation of the segments in their respective principal component without dilution from the other segments. This allows our method to better isolate the segments when compared to sparse PCA.

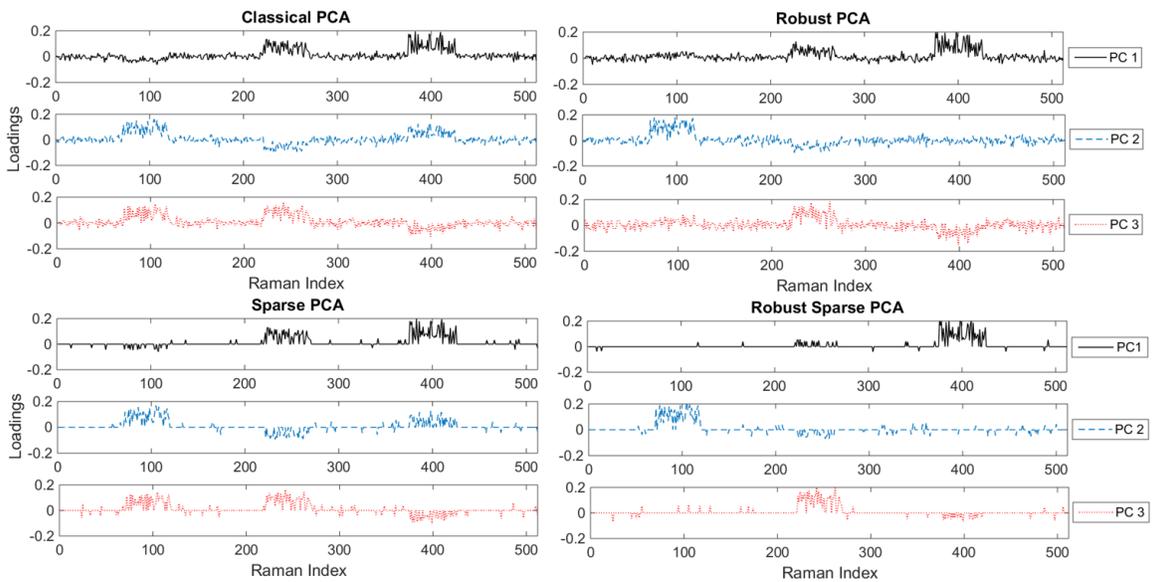

Figure 7 Demonstration of extracted significant principal components of the Raman data

Next, we project the original data on the extracted principal components and test whether the profiles are in control or out of control. Figure 8 shows the plots for the projection of a sample of the original data onto the extracted components (PC scores) for a representative sample. While Figure 9 demonstrates the mean shift in the sparse segments of the out-of-control profiles. Table 2 summarizes the fault detection delay results from 1000 iterations.

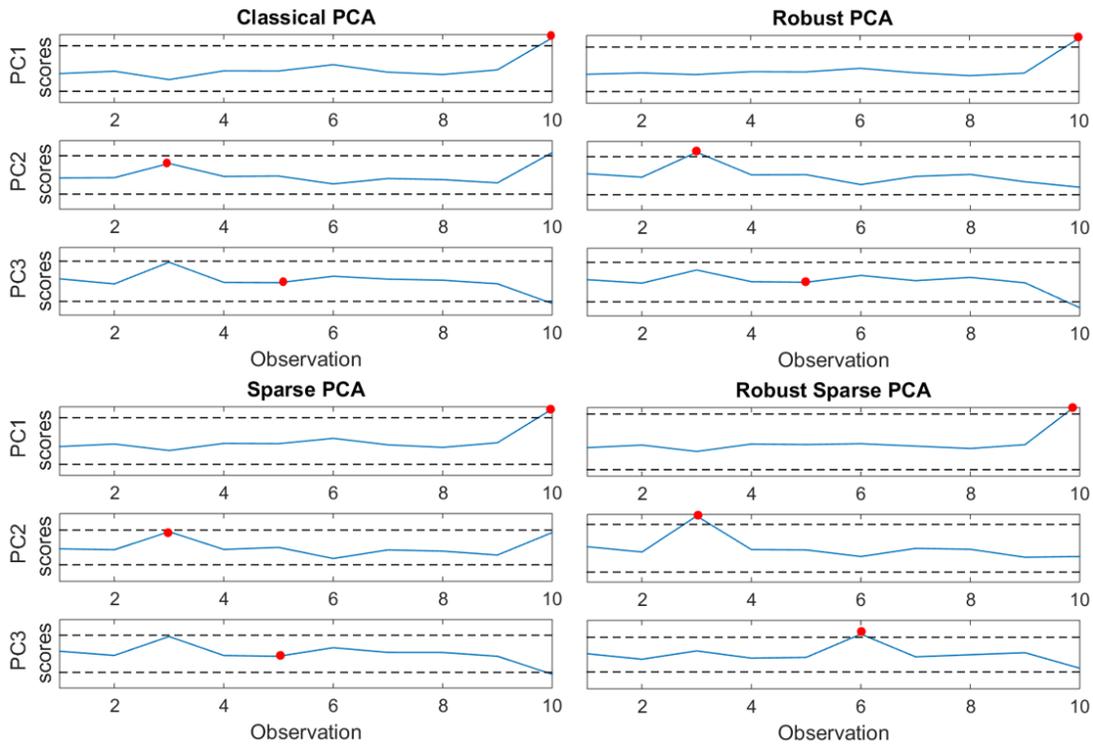

Figure 8 Projection of representative data on the PCs from RSPCA and other benchmarks

Table 2 Detection delay comparison between RSPCA and the other PCA techniques

|  | Defect 1 | Defect 2 | Defect 3 |
|---|---|---|---|
| **Proposed RSPCA** | *4.1(3.4)* | *3.2(3.4)* | *2.5(1.6)* |
| **PCA** | 25.6(10.1) | 17.2(6.2) | 19.7(5.6) |
| **Sparse PCA** | 20.1(5.8) | 18.9(4.3) | 10.6(3.9) |
| **Robust PCA** | 12.1(6.5) | 16(5.1) | 9.6(4.2) |

Note that the defective profiles can be clearly spotted by the projection to the principal component corresponding to the defective segment using RSPCA. These projections are marked by the red dots in Figure 8. The red projections are highly pronounced by our proposed method when compared to the remaining PCA techniques. This is reinforced from the results of Table 2, where the detection delay of the robust sparse PCA is significantly better than its counterparts for defects occurring in any of the regions highlighted in Figure 9.

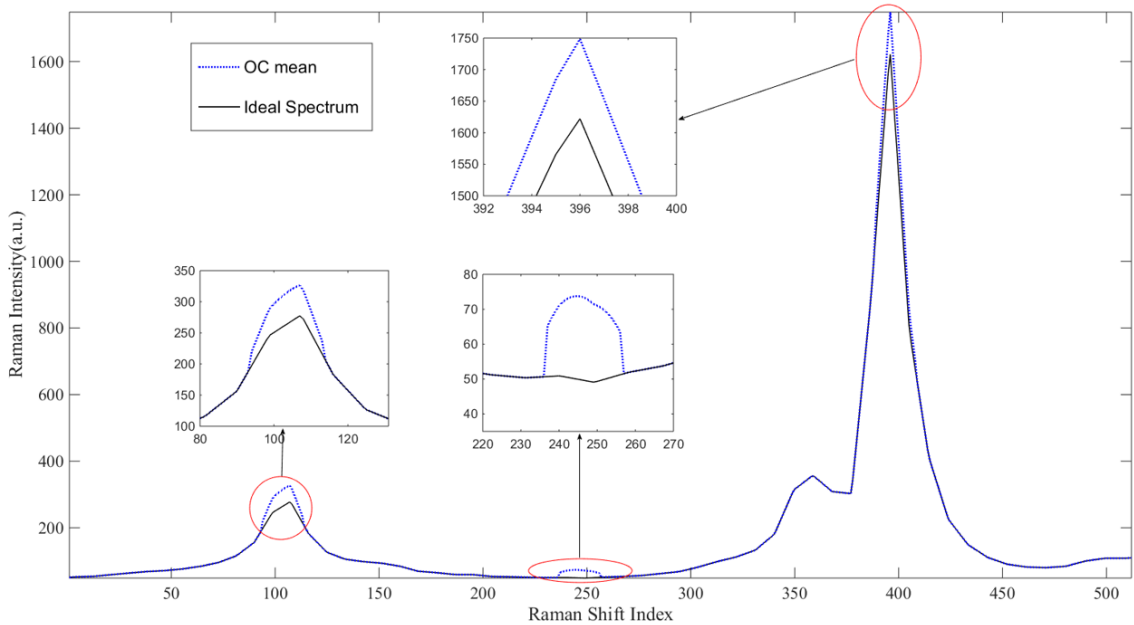

Figure 9 Illustration of the out-of-control shift in the sparse segments of the profiles

# 6. CONCLUSION

Change detection in processes that generate high dimensional data streams has become crucial with the advancement in sensing technologies. This article proposes a novel method that exploits the spatial structure of the data streams while simultaneously reducing the dimension. This is achieved by using a probabilistic model with Laplacian priors to extract robust sparse principal components. This is advantageous because it allows for consistent and flexible modelling of the data streams based on the underlying spatial structure even in the event of noise and outliers.

This research introduces a new feature extraction tool deal with high dimensional data streams for the purpose of process monitoring. It can be used as a feature selection methodology that isolates the streams and capture the general structure, such that they can be monitored without being affected by the noise of insignificant streams or outliers.

We evaluated our RSPCA method against other benchmark PCA based techniques. The results from both the simulation and the case study demonstrate the effectiveness of our proposed procedure in dealing with data with sparse irregularities and outliers. The case study of Raman spectroscopy for bucky-paper manufacturing highlights the capability of the method to isolate sparse segments from high dimensional profiles while mitigating the effect of outliers.

Our proposed method mainly relies on robustly obtaining the sparse principal components based on the structure of the observation variables, and as is with regular PCA, these components are linear combinations of the original data. Kernel PCA (Schölkopf et al. 1997) can accommodate non-linear structures that may be embedded into the original data. Extensions to nonlinear relations were not discussed in this paper but are interesting to explore in their own right in future research.

# REFERENCES

Archambeau, Cédric, and Francis R Bach. 2009. Sparse probabilistic projections. Paper read at Advances in neural information processing systems.

Archambeau, Cédric, Nicolas Delannay, and Michel Verleysen. 2006. Robust probabilistic projections. Paper read at Proceedings of the 23rd International conference on machine learning.

Archambeau, Cédric, Nicolas Delannay, and Michel Verleysen. 2008. "Mixtures of robust probabilistic principal component analyzers." *Neurocomputing* no. 71 (7-9):1274-1282.

Cadima, Jorge, and Ian T Jolliffe. 1995. "Loading and correlations in the interpretation of principle compenents." *Journal of Applied Statistics* no. 22 (2):203-214.

Candès, Emmanuel J, Xiaodong Li, Yi Ma, and John Wright. 2011. "Robust principal component analysis?" *Journal of the ACM (JACM)* no. 58 (3):11.

Chen, Tao, Elaine Martin, and Gary Montague. 2009. "Robust probabilistic PCA with missing data and contribution analysis for outlier detection." *Computational Statistics & Data Analysis* no. 53 (10):3706-3716.

Cheng, Qunfeng, Ben Wang, Chuck Zhang, and Zhiyong Liang. 2010. "Functionalized Carbon‐Nanotube Sheet/Bismaleimide Nanocomposites: Mechanical and Electrical Performance Beyond Carbon‐Fiber Composites." *Small* no. 6 (6):763-767.

Croux, Christophe, Peter Filzmoser, and Heinrich Fritz. 2013. "Robust sparse principal component analysis." *Technometrics* no. 55 (2):202-214.

Croux, Christophe, and Anne Ruiz-Gazen. 2005. "High breakdown estimators for principal components: the projection-pursuit approach revisited." *Journal of Multivariate Analysis* no. 95 (1):206-226.

d'Aspremont, Alexandre, Francis Bach, and Laurent El Ghaoui. 2008. "Optimal solutions for sparse principal component analysis." *Journal of Machine Learning Research* no. 9 (Jul):1269-1294.

De La Torre, Fernando, and Michael J Black. 2003. "A framework for robust subspace learning." *International Journal of Computer Vision* no. 54 (1):117-142.

Ding, Xinghao, Lihan He, and Lawrence Carin. 2011. "Bayesian robust principal component analysis." *IEEE Transactions on Image Processing* no. 20 (12):3419-3430.

Gao, Junbin. 2008. "Robust L1 principal component analysis and its Bayesian variational inference." *Neural computation* no. 20 (2):555-572.

Ge, Zhiqiang, and Zhihuan Song. 2011. "Robust monitoring and fault reconstruction based on variational inference component analysis." *Journal of Process Control* no. 21 (4):462-474.

Guan, Yue, and Jennifer G Dy. 2009. Sparse Probabilistic Principal Component Analysis. Paper read at AISTATS.

Han, Ningning, Yumeng Song, and Zhanjie Song. 2017. "Bayesian robust principal component analysis with structured sparse component." *Computational Statistics & Data Analysis* no. 109:144-158.

Hubert, Mia, Tom Reynkens, Eric Schmitt, and Tim Verdonck. 2016. "Sparse PCA for high-dimensional data with outliers." *Technometrics* no. 58 (4):424-434.

Hubert, Mia, Peter J Rousseeuw, and Karlien Vanden Branden. 2005. "ROBPCA: a new approach to robust principal component analysis." *Technometrics* no. 47 (1):64-79.

Jolliffe, Ian. 2011. "Principal component analysis." In *International encyclopedia of statistical science*, 1094-1096. Springer.

Jolliffe, Ian T, Nickolay T Trendafilov, and Mudassir Uddin. 2003. "A modified principal component technique based on the LASSO." *Journal of computational and Graphical Statistics* no. 12 (3):531-547.

Kim, Dongsoon, and In-Beum Lee. 2003. "Process monitoring based on probabilistic PCA." *Chemometrics and intelligent laboratory systems* no. 67 (2):109-123.

Li, Guoying, and Zhonglian Chen. 1985. "Projection-pursuit approach to robust dispersion matrices and principal components: primary theory and Monte Carlo." *Journal of the American Statistical Association* no. 80 (391):759-766.

Ma, Zongming. 2013. "Sparse principal component analysis and iterative thresholding." *The Annals of Statistics* no. 41 (2):772-801.

Mason, Robert L, Nola D Tracy, and John C Young. 1995. "Decomposition of T2 for multivariate control chart interpretation." *Journal of quality technology* no. 27 (2):99-1108.

Schölkopf, Bernhard, Alexander Smola, and Klaus-Robert Müller. 1997. Kernel principal component analysis. Paper read at International Conference on Artificial Neural Networks.

Tibshirani, Robert. 1996. "Regression shrinkage and selection via the lasso." *Journal of the Royal Statistical Society. Series B (Methodological)*:267-288.

Tipping, Michael E, and Christopher M Bishop. 1999a. "Mixtures of probabilistic principal component analyzers." *Neural computation* no. 11 (2):443-482.

Tipping, Michael E, and Christopher M Bishop. 1999b. "Probabilistic principal component analysis." *Journal of the Royal Statistical Society: Series B (Statistical Methodology)* no. 61 (3):611-622.

Vidal, René, Yi Ma, and S Shankar Sastry. 2016. "Robust Principal Component Analysis." In *Generalized Principal Component Analysis*, 63-122. Springer.

Wang, Chuan, Hsiao-Chun Wu, and Jose C Principe. 1996. Cost function for robust estimation of PCA. Paper read at Aerospace/Defense Sensing and Controls.

Yue, Xiaowei, Kan Wang, Hao Yan, Jin Gyu Park, Zhiyong Liang, Chuck Zhang, Ben Wang, and Jianjun Shi. 2017. "Generalized wavelet shrinkage of inline Raman spectroscopy for quality monitoring of continuous manufacturing of carbon nanotube buckypaper." *IEEE Transactions on Automation Science and Engineering* no. 14 (1):196-207.

Yue, Xiaowei, Hao Yan, Jin Gyu Park, Zhiyong Liang, and Jianjun Shi. 2018. "A Wavelet-Based Penalized Mixed-Effects Decomposition for Multichannel Profile Detection of In-Line Raman Spectroscopy." *IEEE Transactions on Automation Science and Engineering, 15(3), pp.1258-1271*.

Zeng, Jing, Kangling Liu, Weiping Huang, and Jun Liang. 2017. "Sparse probabilistic principal component analysis model for plant-wide process monitoring." *Korean Journal of Chemical Engineering* no. 34 (8):2135-2146.

Zhang, Zhenyue, Hongyuan Zha, and Horst Simon. 2002. "Low-rank approximations with sparse factors I: Basic algorithms and error analysis." *SIAM Journal on Matrix Analysis and Applications* no. 23 (3):706-727.

Zhang, Zhenyue, Hongyuan Zha, and Horst Simon. 2004. "Low-rank approximations with sparse factors II: Penalized methods with discrete Newton-like iterations." *SIAM journal on matrix analysis and applications* no. 25 (4):901-920.

Zhu, Jinlin, Zhiqiang Ge, and Zhihuan Song. 2014. "Robust modeling of mixture probabilistic principal component analysis and process monitoring application." *AIChE journal* no. 60 (6):2143-2157.

Zou, Hui, Trevor Hastie, and Robert Tibshirani. 2006. "Sparse principal component analysis." *Journal of computational and graphical statistics* no. 15 (2):265-286.